# Subtractor-Based CNN Inference Accelerator

Victor Gao[1], Issam Hammad[2], *Member, IEEE*, Kamal El-Sankary[1], *Senior Member, IEEE,* and Jason Gu[1], *Senior Member, IEEE*

[1]The Department of Electrical and Computer Engineering, Dalhousie University, Halifax, NS, Canada
[2]The Department of Engineering Mathematics and Internetworking, Dalhousie University, Halifax, NS, Canada

*Abstract*— This paper presents a novel method to boost the performance of CNN inference accelerators by utilizing subtractors. The proposed CNN preprocessing accelerator relies on sorting, grouping, and rounding the weights to create combinations that allow for the replacement of one multiplication operation and addition operation by a single subtraction operation when applying convolution during inference. Given the high cost of multiplication in terms of power and area, replacing it with subtraction allows for a performance boost by reducing power and area. The proposed method allows for controlling the trade-off between performance gains and accuracy loss through increasing or decreasing the usage of subtractors. With a rounding size of 0.05 and by utilizing LeNet-5 with the MNIST dataset, the proposed design can achieve 32.03% power savings and a 24.59% reduction in area at the cost of only 0.1% in terms of accuracy loss.

*Keywords*— AI Accelerator, Convolutional Neural Networks (CNN), Deep Learning, Weight Approximation, Weight Sorting,

## I. Introduction

Deep learning using convolutional neural networks (CNNs) is now widely employed in various computer vision applications. CNNs have achieved classification accuracy levels that surpass those of humans [1-3]. These networks find applications in diverse industries and disciplines, including real-time image classification [4], human action recognition [5], brain tumor detection [6], and the detection of structural damage in nuclear reactors [7]. However, achieving more accurate predictions often requires larger CNN networks, which demand higher computational power. Therefore, introducing computational methods that can reduce CNN complexity and, consequently, overall power consumption is essential, especially for embedded systems that rely on batteries. This is particularly crucial for optimizing the performance of convolutional layers [8]. Such energy reduction is necessary for AI accelerators used in mobile devices, AV/AR devices, drones, and other embedded systems. Previously, several methods have been proposed to reduce CNN computation complexity. These methods include parameter pruning [9-12], weights sparsity utilization [13], approximate computing, which involves approximate multipliers [14-19], and various weight quantization methods [20-22]. These approaches manipulate network parameters to reduce power, area, and delay, providing energy-efficient solutions for CNN computations.

This paper proposes an energy-efficient design for CNN inference by introducing a method that can replace part of the required multiplications and additions with subtractions during the inference stage. Given the high cost of multiplication in terms of energy and the much lower costs of subtraction, this substitution allows for a substantial reduction in the required power and area of the system. Figure 1 presents the inference computational time percentage for each layer in AlexNet [23]. As shown in the figure, the convolutional layers use around 90% of the total processing time in both a CPU and GPU [8] setting. Therefore, any performance enhancements to the convolutional layers will have a major impact on the system as a whole. The proposed design method focuses on pre-trained networks for utilization during inference. The method starts with a trained model, then the weights are extracted to find combinations that enable subtraction. During inference, the modified convolution unit is utilized to handle the modified weights. The paper summarizes potential performance enhancements that can be achieved for various rounding sizes. The paper is organized as follows: Section 2 presents the research background and the motivation, Section 3 describes the implementation details, Section 4 presents the simulation results, and Section 5 presents the research conclusion.

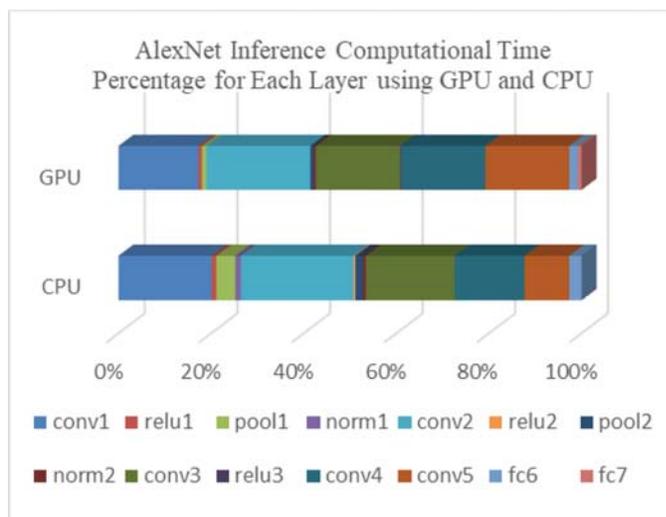

*Fig. 1 AlexNet inference computational time percentage for each layer*

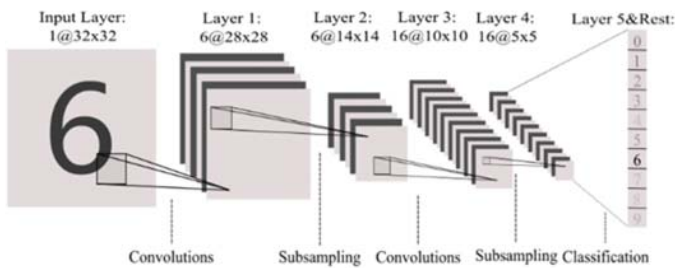

*Fig. 2 LetNet-5 architecture*

## II. BACKGROUND AND MOTIVATION

To demonstrate the performance enhancements of the proposed method, the popular LeNet-5 CNN network was utilized [19]. The architecture of LeNet-5 is shown in Fig.2. As can be seen from the figure, in layer 1, the input data is represented by a single channel 32 x 32 pixels image for a handwritten number, and the output is a Softmax function with ten nodes representing the digits from zero to nine. To explore the utilization of the subtractors option, an analysis of the distribution of the weight was performed. Fig. 3 illustrates the weights of the third convolutional layer in LeNet-5, while Fig. 4 shows the histogram for the distribution of the weights. As can be seen from the figure, the distribution allows for finding opposite (negative and positive) pairs weights that can be combined; the proposed method exploits this property by utilizing subtractions to replace additions and multiplications, as will be presented in the next section.

## III. IMPLEMENTATIONS

This section provides an overview of the proposed method, which is summarized in Figure 5. The proposed implementation relies on utilizing two blocks: a weight preprocessor and a modified convolution unit. The weights preprocessing occurs once before deploying the weights for inference. The preprocessor prepares the weights for use by the modified convolution unit during the inference stage. The first preprocessing step involves sorting and splitting the weights into two lists: one for positive weights and one for negative weights. In the second step, the preprocessor identifies all possible combinations based on the selected rounding step and creates a list of combined weights. Finally, the preprocessor combines all three lists and replaces the original weights in the CNN model with the modified weights for inference. During the inference stage, the modified convolution unit handles the combined and uncombined weights separately. The combined weights rely on the subtraction operation to replace one addition and multiplication, while the uncombined weights will use regular addition and multiplication. More details about the preprocessing step are presented in subsection A, while subsection B presents more details about the modified convolution unit.

### A. Preprocessing of the Weights by Sorting and Approximation

Preprocessing of the weights starts by sorting them, then finding combinations to merge, as shown in Figure 6. Initially, the weights are sorted in ascending order and split into two lists: one for positive weights and one for negative weights. The simulation of this process was performed using Numpy [24]. The preprocessor in NumPy saves the original positions of the weights during the sorting process, using a flag to indicate the status of each weight as processed, combined, or not combined. After sorting, the weights are combined based on a specified rounding size, resulting in a new list that contains all the combined weights from the positive and negative weight lists. All three lists are then merged and spliced to have all the combined weights at the top, while the rest of the uncombined weights are at the bottom, as depicted in Figure 6.

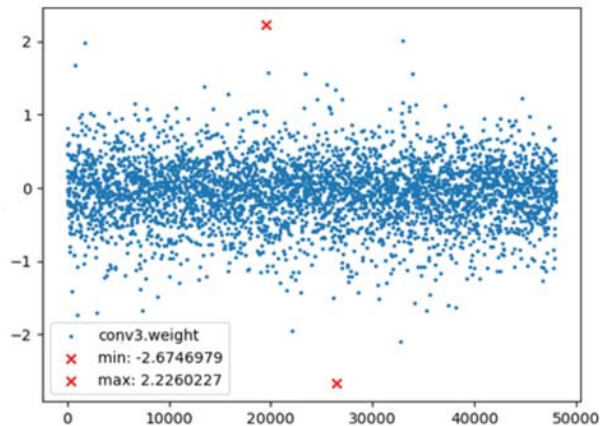

*Fig. 3 Weight distribution in the third convolutional layer*

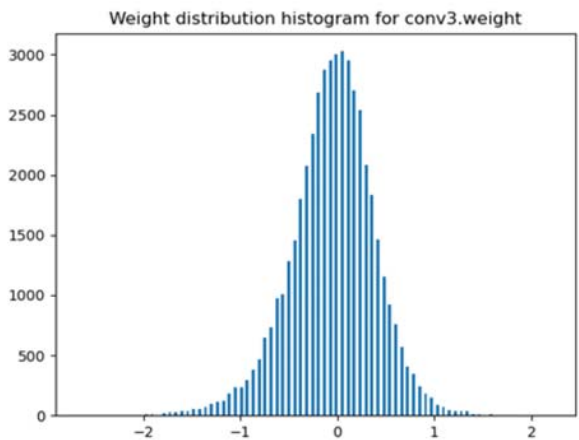

*Fig. 4 Histogram of weight distribution*

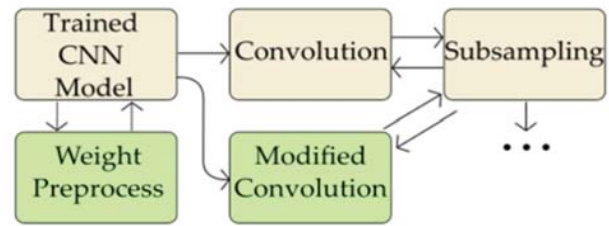

*Fig. 5 Structure of the proposed accelerator*

This work is funded by the Natural Sciences and Engineering Research Council of Canada (NSERC)

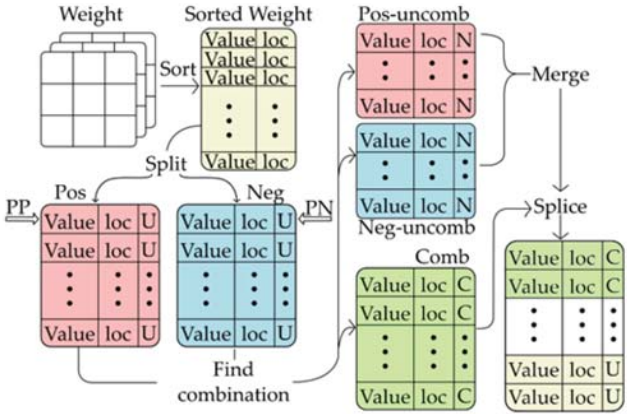

*Fig. 6 Details of the weight sorting and grouping*

### B. Combing the weights for convolution

The process of combing the weights which was presented in section 3 allows for the utilization of one subtraction as a replacement for one multiplication and one addition operation as illustrated in (1).

$$I_1 \times K_a + I_2 \times K_b = K_a \times (I_1 - I_2) \quad if\ K_a = -K_b \quad (1)$$

The sorted weights will rely on the extracted position value, which is generated during preprocessing part during the inference stage. As for the uncombined weights, they will simply use the regular CNN inference multiplications and additions.

```
Algorithm 1 Find combinations
Input: Pos, Neg, PP, PN, rounding   ▷ List of sorted
   positive and negative weights, PP and PN are the pointers
   pointing to sorted positive and negative weight lists
Output: Comb, Pos − uncomb, Neg − uncomb   ▷ List
   of found combinations weights, and rest weights stay in
   original list
 1: idx ← 0
 2: comb ← empty          ▷ Initialize empty list
 3: while PP and PN exists do
 4:    if PP.val ≥ |PN.val| + rounding then  ▷ Negative
       weights too small
 5:       PN.U ← N   ▷ Assign N as no combination to
          current weight status
 6:       Inc PN              ▷ Point to next weight
 7:    else if PP.val ≤ |PN.val| − rounding then
 8:       PP.U ← N
 9:       Inc PP
10:    else
11:       PP.U ← C   ▷ Assign C as combination exists
12:       PN.U ← C
13:       comb[idx] ← PP  ▷ Store current weight element
          to comb list
14:       comb[idx + 1] ← PN
15:       idx ← idx + 2
16:       Delete PN, PP
17:       Inc PP, PN
18:    end if
19: end while
```

## IV. RESULTS

The proposed method's performance enhancements were evaluated in terms of power and area using a frequency of 1GHz and the Design Compiler from Synopsys with TSMC 65nm technology. All tested mathematical operations, including multiplication, subtraction, and addition, adhered to the IEEE 758 design standard. The software implementation of the CNN network was tested using LeNet-5 with MNIST data, employing Numpy and Pytorch [25]. Table 1 illustrates the number of additions, subtractions, and multiplications for different rounding sizes. The table demonstrates that increasing the rounding size results in a higher number of subtractions while reducing both additions and multiplications. A larger step size leads to a reduction in the total number of operations. Figure 7 illustrates a bar chart for the distribution of mathematical operations for various rounding sizes..

TABLE I.   NUMBER OF ADDITION, SUBTRACTION, AND MULTIPLICATION WITH DIFFERENT ROUNDING SIZES FOR LENET-5

| Rounding Size | Additions | Subtractions | Multiplications | Total |
|---|---|---|---|---|
| 0 | 405600 | 0 | 405600 | 811200 |
| 0.0001 | 399372 | 6228 | 399372 | 804972 |
| 0.005 | 313545 | 92055 | 313545 | 719145 |
| 0.01 | 288887 | 116713 | 288887 | 694487 |
| 0.015 | 276692 | 128908 | 276692 | 682292 |
| 0.02 | 265480 | 140120 | 265480 | 671080 |
| 0.025 | 259789 | 145811 | 259789 | 665389 |
| 0.05 | 242153 | 163447 | 242153 | 647753 |
| 0.1 | 233698 | 171902 | 233698 | 639298 |
| 0.15 | 228752 | 176848 | 228752 | 634352 |
| 0.2 | 225988 | 179612 | 225988 | 631588 |
| 0.25 | 223630 | 181970 | 223630 | 629230 |
| 0.3 | 222742 | 182858 | 222742 | 628342 |

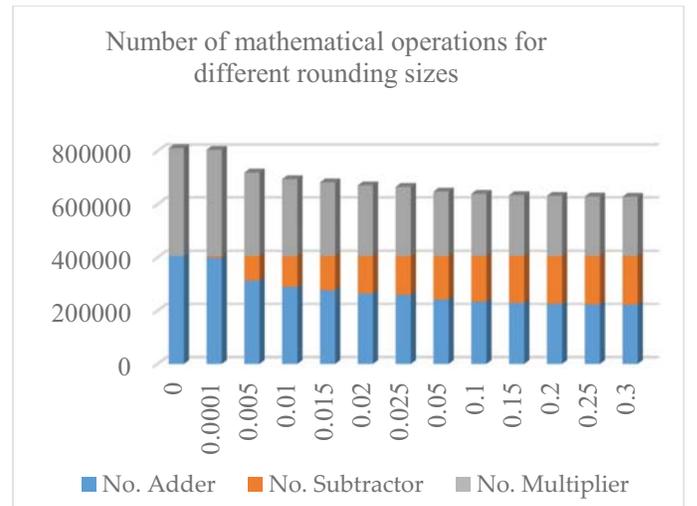

*Fig. 7 Mathematical operations distribution for different rounding sizes*

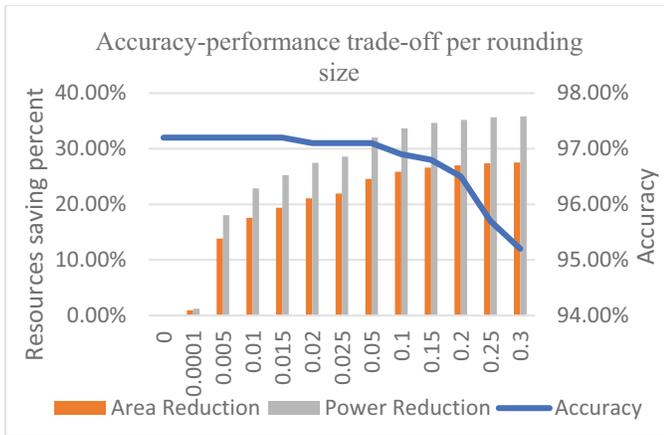

*Fig. 8 Accuracy-performance trade-off per rounding size*

Figure 8 shows the relationship between rounding size, power, area, and accuracy. The percentage on the left represents the percentage of power and area savings, while the percentage on the right represents the CNN classification accuracy, which drops with a higher rounding size. As shown in Figure 8, the accuracy drops dramatically after a step size of 0.05. Thus, there is a trade-off between power, area saving, and accuracy. With a step size of 0.05, the power can be reduced by 32.03%, and the area can be reduced by 24.59%, resulting in an accuracy loss of only 0.1%.

V. CONCLUSIONS

This paper presented a novel method to reduce the power and area of CNN inference accelerators by replacing one multiplication and one addition operation with one subtraction operation. The proposed method allows for a significant performance improvement in terms of power and area saving with minimal accuracy loss. The paper presented the trade-off that can be achieved between performance enhancement and accuracy loss based on the selected rounding size. As shown in the paper, with a rounding size of 0.05, a power reduction of 32.03% and an area reduction of 24.59% can be achieved with only a 0.1% accuracy loss. The design allows for adjusting the trade-off between gained performance enhancements and the cost in accuracy loss.